\documentclass[twocolumn,showpacs,preprintnumbers,amsmath,amssymb]{revtex4}

\usepackage{graphicx}

\usepackage{dcolumn}  % Align table columns on decimal point
\usepackage{bm}       % bold math

\newcommand{\be}{\begin{eqnarray}}
\newcommand{\ee}{\end{eqnarray}}

\newcommand{\beq}{\begin{equation}}
\newcommand{\eeq}{\end{equation}}

\begin{document}

\title{Relativistic and spin effects in elastic backward p-d
 scattering}
\author{A.P.Ierusalimov$^{a}$}
\author{G.I.Lykasov$^{a}$}
\email{lykasov@jinr.ru}
\author{M.Viviani$^{b}$}
\affiliation{$^{a}$JINR, Dubna, Moscow region, 141980, Russia, \\ 
$^{b}$INFN, Sezione di Pisa, Largo Bruno Pontecorvo, I-56127, 
Pisa, Italy}
\date{\today}

\begin{abstract}
The elastic backward proton-deuteron scattering is analyzed including 
relativistic effects in the deuteron and the mechanism of this reaction
which includes the graphs corresponding to the emission, 
rescattering and absorption of the virtual pion by a deuteron nucleon 
in addition to the one-nucleon exchange graph. It allows one to obtain 
a rather satisfactory description of all the experimental data on the 
differential cross section, tensor analyzing power of the deuteron and 
transfer polarization in this reaction.
\end{abstract}

\maketitle

%%%%%%%%%%%%%%%%%%%%%%%%%%%%%%%%%%%%%%%%%%%%%%%%%%%%%%%%%%%%%%%%%%
\section{Introduction}
As is well known, the study of polarization phenomena in hadron and hadron-nucleus 
interactions gives more detailed information on dynamics of their interactions
and the structure of colliding particles. The quark structure and relativistic
effects of light nuclei, in particular, deuterons, is one of important problems in nuclear 
physics at intermediate and high energies. The theoretical and experimental
study of reactions like the elastic 
$e-d$ 
\cite{Arnold:1981} 
and $p-d$ 
\cite{Arvieux:1984,Keister:1981} 
scattering, deuteron break-up reactions induced 
by electrons or protons 
\cite{Rekalo,Rekalo1} 
and the deuteron stripping processes on protons and 
nuclei at intermediate and high energies 
\cite{Perdrisat:1987,DL:1990},
can allow us to find out new information on the deuteron structure at short 
distances. The elastic backward proton-deuteron scattering has been 
experimentally and theoretically studied in Saclay 
\cite{Arvieux:1984}, Dubna 
and at JLab (USA) 
\cite{Azhgirei,Azhgirei1,Azhgirei2,Punjabi}. 
 Usually these processes are analyzed within a simple
impulse approximation.
Up to now all these data have not been described within the one-nucleon exchange 
model (ONE) including even the relativistic effects in the deuteron 
\cite{KermKissl:1969,IL:2001}.   
In this paper we analyze the elastic backward proton-deuteron scattering within 
the relativistic approach including the ONE and the high order graphs 
corresponding to the emission, rescattering and absorption of the virtual pion 
by a nucleon of deuteron 
\cite{ILV:2007}. 

%%%%%%%%%%%%%%%%%%%%%%%%%%%%%%%%%%%%%%%%%%%%%%%%%%%%%%%%%%%%%%%%%%
\section{Light cone dynamics for $dp\rightarrow pd$}
\subsection{The leading order diagrams}
Let us analyze the elastic $dp\rightarrow pd$ scattering within the Weinberg
diagrammatic technique in the infinite-momentum frame (IMF) 
\cite{Weinberg:1966,Brodsky:1973}.
The four-momentum of the fast deuteron $P_d$ and its nucleons $k_1$ and $k_2$ 
have the following components in the IMF:
\be
P_d\left(P+\frac{m_d^2}{2P}, 0, P\right) \\
\nonumber
k_1\left(xP+\frac{m_t^2}{2xP}, {\bf k}_t, xP\right) \\
\nonumber
k_2\left((1-x)P+\frac{m_t^2}{2(1-x)P}, -{\bf k}_t, (1-x)P\right)~,
\label{def:IMF}
\ee 
where $P$ is the magnitude of the three-momentum of the incident particle,
in particular, deuteron;
$x=(E(k)+k_z)/(E_d(p_d)+p_{dz})$ is the light cone variable, 
$E(k)=\sqrt{{\bf k}^2+m^2_N}$ and 
$E_d(p_d)=\sqrt{{\bf p}^2_d+m^2_N}$ are the total energies of the nucleon inside 
deuteron and of the deuteron, respectively; ${\bf k},{\bf p}_d$ are three-momenta of 
this nucleon and deuteron; $k_z,p_{dz}$ are their longitudinal components;
$m_N$ is the nucleon mass.

The first order diagrams or the one-nucleon exchange graphs are presented in Fig.1.
In the general relativistic case the deuteron vertex $dpn$ does not reduce merely 
to dissociation of the deuteron into two nucleons; it may also include the 
annihilation ${\bar N}d\rightarrow N$ and therefore, the deuteron decay vertex 
cannot always be reduced to an ordinary deuteron wave function whose square is 
a probability of finding a nucleon in the deuteron with a definite momentum. As 
is well known, Feynman graph of $n$th order is equivalent to $n!$ time ordered
graphs of the old perturbative theory (OPT). If $dp$ processes are analyzed within
the IMF, then many graphs in the old perturbative theory make a contribution
of order $O(1/P)$ 
\cite{Weinberg:1966,Brodsky:1973}.
 There remain only the diagrams that correspond 
to the dissociation of the deuteron into two nucleons.
As an example, Fig.1 shows the Feynman diagram of the $dp\rightarrow pd$ process 
(Fig.1a) and two equivalent diagrams of the OPT, ordered in the time $t$ 
(Figs.1(b,c)). The graph of Fig.1a corresponds to the deuteron dissociation and 
the graph of Fig.1b is the so called $z$-diagram corresponding to the 
${\bar N}d\rightarrow N$ annihilation.

Therefore, one can introduce the concept of a deuteron wave function (d.w.f.) 
with the usual probability interpretation. In this case the d.w.f. $\Psi$ depends
on the following relativistic invariant variable 
\cite{FS:1981,SF:1980}:
\begin{equation}
 k^2~=~\frac{m_t^2}{4x(1-x)}~-~m_N^2.
\label{def:ksqure}
\end{equation}    
In \cite{SF:1980} it was shown that the variable $k^2$ is proportional to the 
difference of the initial and final energies in the $d\rightarrow pn$ dissociation
vertex of Fig.1b.

Note that in each vertex of the OPT graph (Figs.1(b,c)) the three-momentum is 
conserved but the energy is not, although the energy and the three-momentum is 
conserved for the complete reaction. All the particles, including those in the 
intermediate state, are on the mass shell. In the Feynman-diagram technique the 
four-momentum is conserved at each vertex of the diagram (Fig.1a), but the 
intermediate particle with four-momentum $k_N$ is off the mass shell, e.g., 
$k_N^2\neq m^2$.

%%%-----------  Fig.1 ----------------
\begin{figure}[htb]
\includegraphics[width=0.48\textwidth]{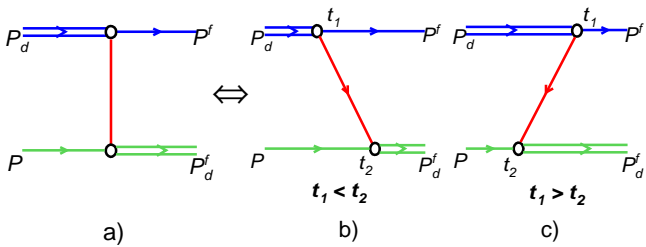}
\caption[Fig.1]{The Feynman graph corresponding to the one-nucleon exchange 
graph (a) for the process $d p\rightarrow p d$ (a) and its equivalent to two
time-ordered diagrams within the OPT (b,c) .}
\label{Fig.1} 
\end{figure}

In such approach 
\cite{FS:1981,Kobushkin:1982},
 the d.w.f. $\Psi_d$ is related to the nonrelativistic d.w.f. $\Phi^{n.r.}_d$,
that depends on the relativistic invariant variable $k^2$ given by Eq.
(\ref{def:ksqure}):
\begin{equation}
\Psi_d(x,k_t)~=~\left(\frac{m^2_t}{4x(1-x)}\right)^{1/4}\Phi^{n.r.}_d(k^2)~,
\label{def:psid}
\end{equation}
with the following normalization equation:
\begin{equation}
\frac{1}{2}\int_0^1\frac{dx}{x(1-x)}\int\mid\Psi_d(x,k_t)\mid^2d^2k_t~=~1~.
\label{def:normpsi}
\end{equation}  

There are also several covariant approaches to construct the relativistic d.w.f. 
For example, one of them \cite{BuckGross:1979,Tokarev:1991} is based on the 
assumption that one nucleon inside deuteron is on its mass shell ($k_1^2=m_N^2$)
while the other one is off-mass-shell ($k_2^2-m_N^2<0$ or $k_2^2-m_N^2>0$).
In that case additionally to the $S$- and $D$-wave functions ($u,w$) in the d.w.f. 
two components also appear: the triplet $P$-state wave function ($v_t$) and
the singlet $P$-state wave function ($v_s$). Actually, as is shown in 
\cite{BuckGross:1979}, 
the total probability for the $P$-wave components in the d.w.f. ($W_p$) is too 
small, it is less than $1\%$. Recently it has been shown 
\cite{IL:2001} 
that within the relativistic one-nucleon exchange model (RONE) proposed in 
\cite{BuckGross:1979} 
one can not describe all the polarization experimental data for the elastic
backward $p-d$ scattering even by increasing the parameter $W_p$ till several
percent. 

Another interesting covariant approach within the light cone dynamics to
construct the relativistic d.w.f. \cite{Karmanov,Karmanov1} is based on the
three-dimensional formalism for the quantum field theory. Within this approach two
nucleons inside the deuteron are mass-shell, however, the four-vector $\omega$ 
determining the light cone surface is introduced to satisfy the three-momentum 
and energy conservation by the deuteron break-up. The relativistic d.w.f. 
constructed within this approach depends on the relativistic invariant variable
$k^2$ and direction vector ${\bf n}$ of the IMF. For example, choosing
${\bf n}$ in the opposite direction to the deuteron moving in the IMF, one gets
the same dependence of $\Psi_d$ on $k^2$ given by Eq.
(\ref{def:psid})
like in 
\cite{FS:1981,Kobushkin:1982}.            
Other approaches to get the relativistic d.w.f. can be found, for example, in 
\cite{Keister:1981,Tjon:1982} 
and 
\cite{Lev}. 

The  amplitude for the elastic backward $dp$ scattering within the impulse
approximation of the OPT (Fig.1b) in the LCD has the following form 
\cite{GL:1993,Yudin:2000}:
\begin{equation}
{\cal F}^{(1)}_{LCD}=\sqrt{3}\frac{M_d^2-m_N^2/(x(1-x))}{1-x}\Psi_d^2(k^2)~,
\label{def:FLCONE}
\end{equation}
On the other hand, the amplitude corresponding to the one-nucleon exchange Feynman
graph (Fig.1a) can be presented in the following form 
%\cite{Rekalo,Sitnik:1994,IL:2001}:
\cite{Rekalo,Sitnik:1994,Sitnik96,Sitnik98,IL:2001}:
\begin{equation}
{\cal F}^{RONE}=8\sqrt{3} m_N(m_N^2-u)\Psi_d^2(k^2)~,
\label{de:FRONE}
\end{equation}
where 
$u$ is the square of momentum transfer from initial deuteron to final proton;
$k^2$ can be also written in the following form: 
$k^2=\frac{1}{4}s_{12}-m^2_N$; $s_{12}=(k_1+k_2)^2$; $k_1,k_2$ are the four-momenta
of neutron and proton in the deuteron. Unfortunately, the ONE and the RONE do not
allow a satisfactory description of all the observables at the kinetic energy of
backward scattered protons $T_p>0.6$ GeV 
\cite{IL:2001}.  

%%%%%%%%%%%%%%%%%%%%%%%%%%%%%%%%%%%%%%%%%%%%%%%%%%%%%%%%%%%%%%%%%%%%%%%%%%%%%%%%%%%%
\subsection{Next to leading order diagrams}

As was shown in 
\cite{Wilkin:1969,Nakamura:1985},
the contribution of the high-order graphs in the $p-d$ backward elastic scattering
corresponding to the emission, scattering and absorption of the virtual pion by a
deuteron nucleon, can be sizable at initial energies corresponding to possible 
production of the $\Delta$-isobar at the $\pi-N$ vertex, see Fig.2(a,b). In 
\cite{Kaptari:1998} this process was analyzed within the
Bethe-Saltpeter approach using the impulse approximation, however, the one-pion
exchange contribution in the intermediate state was also included. The
contribution of the $\Delta$-isobar exchange graph to the elastic $p-d$
scattering was studied in 
\cite{Uzikov,Uzikov1}.

All these models reproduce the gross features of the backward cross section
and describe the experimental data rather well, however, there is a
difficulty to describe both the cross section and the polarization observables
like the tensor analyzing power $T_{20}$ of deuteron and the transfer
polarization $\kappa_0$ within all these approaches. In
\cite{Vasan:1973} it is stressed that the energy dependence of $T_{20}$ should
be sensitive  to the microscopic structure of the model. 
%%%-----------  Fig.2 - 2 pos.---------  
\begin{figure}[htb]
\includegraphics[width=0.45\textwidth]{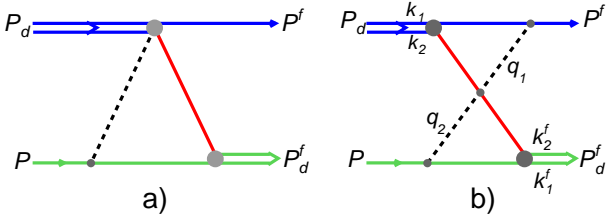}
\caption[Fig.2]{The triangle Feynman graph with one-pion
exchange for the process
$d p\rightarrow p d$, and its equivalent graph (b).} 
\label{Fig.2}
\end{figure}
%%%-----------  Fig.3 - 2 pos.---------
\begin{figure}[htb]
\includegraphics[width=0.45\textwidth]{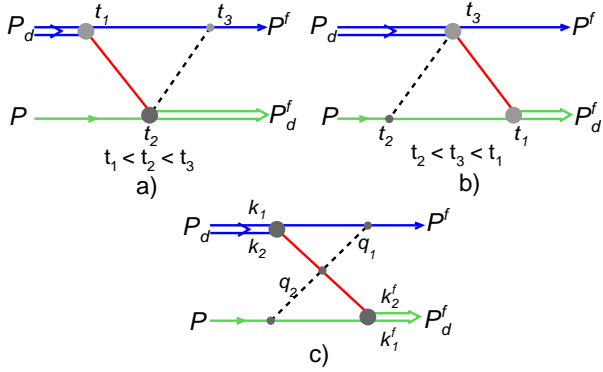}
\caption[Fig.3]{The time ordered graphs with one-pion exchange
within the OPT (a,b) and their equivalent graph (c)
for the process $d p\rightarrow p d$.} 
\label{Fig.3}
\end{figure}

For example, the inclusion of the
triangle Feynman graph of Fig.2a in addition to the one-nucleon exchange 
diagram of Fig.1a allows to describe the experimental data on $T_{20}$
at the deuteron momentum $P_d$ less than $4$ GeV$/c$ only 
\cite{Nakamura:1985},
and this calculation does not describe the tail of $T_{20}$ at$P_d\geq 4$ GeV$/c$.

The corrections to the ONE graph of Fig.1a were also analyzed in other papers,
see, for example,
\cite{Uzikov}
and references therein. As was shown in
\cite{DL:1990,DL1:1990,DN:1989}, 
the contribution of the one-pion exchange graphs to the 
deuteron stripping reaction of the type $d+p\rightarrow p+X$, can be also sizable
at the initial energies close to a possible $\Delta$-isobar production in the
intermediate state.
%%%-----------  Fig.2 - 1 pos.---------
%%%-----------  Fig.3 - 1 pos.---------

Let us apply the Weinberg diagram formalism
\cite{Weinberg:1966} 
within the LCD analyzed in 
\cite{GL:1993,DL:1990} for the deuteron stripping reactions $dp\rightarrow pX$ 
to the elastic $d-p$ scattering. As is known, Feynman graph of the $n$th order
is equivalent to $n!$ time-order graphs of the old perturbative theory (OPT). In 
\cite{Brodsky:1973}, it is shown that the time ordered diagrams of the order $n$
at $x>0$, are finite, whereas at $x<0$ they can be suppressed as $1/P^{n-1}$. One
Feynman diagram of the $3$ order presented in Fig.2a is equivalent to $6$ time-
ordered diagrams calculated within the OPT
\cite{Brodsky:1973,GL:1993},
however, only two diagrams presented in Fig.3(a,b) are finite, while the other $4$ 
graphs are suppressed as $1/P^2$ or as $1/P$ when the spin structure of the
vertices is included, therefore they can be neglected at high values of $P$.
Actually, these results were obtained in \cite{Brodsky:1973} for a $\phi^3$
interaction, nevertheless, it can be also applied for $d-p$ reactions, shown in 
\cite{DL:1990,DL1:1990,GL:1993}.
 
The calculation of the graphs of Fig.3(a,b) is equivalent to
the calculation of the diagram in Fig.3c.    
 
The four-momentum of the fast deuteron $P_d$ and its nucleons $k_1$ and $k_2$ are
represented within the IMF in the same forms, as in \cite{DL:1990}, see Eq.(\ref{def:IMF}).
The part of the $d-p$ elastic scattering amplitude corresponding to the graph of
Fig.(1c) within the OPT in the LCD, can be presented in the following form: 
\cite{GL:1993,DL:1990}:
\begin{widetext}
\be
{\cal F}^{(3)}_{LCD}=-\left(\frac{g P}{(2\pi)^3}\right)^2\int\frac{dx dx^\prime 
d^2k_t d^2k^\prime_t}{4\sqrt{E_N(k_1)E_N(k_2)}4\sqrt{E_N(k_1^f)E_N(k_2^f)}
E_\pi(q_1)E_\pi(q_2)} \\
\nonumber
\Psi^+_d(x^\prime, k^\prime_t)\Gamma_N^{(2)}
 G(q_2)F_{\pi N}(q_2^2)f_{\pi N}^{el}(s_1,t_1)F_{\pi N}(q_1^2)G(q_1)\Gamma_N^{(1)}
\Psi_d(x,k_t)~,
\label{def:lctr}
\ee
\end{widetext}
where $x,x^\prime$ are the light cone variables for nucleons inside the initial and
final deuterons, respectively, while ${\bf k}_t,{\bf k}^\prime_t$ are the transverse
momenta of these nucleons; the energy Green functions $G(q_{1,2})$ within the OPT
have the following forms:
\be
G(q_1)=\left(E_N(p^f)-E_N(k_1)-E_\pi(q_1)+i\epsilon\right)^{-1}\\
\nonumber
G(q_2)=\left(E_N(k_1^f)-E_N(p )-E_\pi(q_2)-i\epsilon\right)^{-1}~,
\label{def:GOPT}
\ee
$$E_N(k_{1,2})=\sqrt{{\bf k}_{1,2}^2+m_N^2},$$ 
$$ E_N(k^f_{1,2})=\sqrt{{{\bf k}^f}^2_{1,2}+m_N^2},$$ 
$$E_\pi(q_{1,2})=\sqrt{{\bf q}_{1,2}^2+\mu_\pi^2},$$ 
where ${\bf k}_{1,2}$ and ${\bf k}^f_{1,2}$ are the three-momenta of nucleons inside
the initial and final deuterons respectively; ${\bf q}_{1,2}$ are three-momenta
of the intermediate pion in Fig.(1c); $\mu_\pi$ is the pion bar mass;
$F_\pi(q^2_{1.2})$ is the pion form factor taking into account the virtuality of
the intermediate pion depending on its four-momentum squared $q_{1,2}^2$. The
pion form factor was taken in the monopole form 
$F_\pi=\Lambda_\pi^2/(\Lambda_\pi^2-q_{1,2}^2)$ , where the value of the cut-off
parameter $\Lambda_\pi$ was taken as $\Lambda_\pi=0.7-0.8$ GeV$/c$ also used in
\cite{DL:1990,DL1:1990} 
and enabled us to make a rather satisfactory description of the experimental on the
$dp\rightarrow pX$ reactions. The d.w.f. $\Psi_d$ is related to the nonrelativistic
d.w.f $\Phi^{n.r.}_d$, see Eq.(\ref{def:psid}), that has the following form
\cite{Reid:1968}:
\begin{equation}
\Phi^{n.r.}_d(k^2)~=~\left(u(k^2)~-~\frac{1}{\sqrt{8}}w(k^2){\cal S}_{np}\right)
\chi_{1M}~,
\label{def:Phinr}
\end{equation} 
where $u(k^2)$ and $w(k^2)$ are the $S$- and $D$-waves of the d.w.f., $\chi_{1M}$ 
is the spin triplet wave function, ${\cal S}_{np}=3({\vec\sigma}_n\cdot{\hat{
\vec k}})({\vec\sigma}_p\cdot{\hat{\vec k}})-({\vec\sigma}_n\cdot{\vec \sigma}_p)$;  
$\hat{\vec k}$ is the unit vector of the relative momentum of nucleons in the
deuteron; $f_{\pi N}(s_1,t_1)$ is the amplitude of the elastic $\pi-N$ scattering, 
see Fig.2b; the vertex $\Gamma_N^{(1)}={\bar u}(p^f)\gamma_5 u(k_1)=\xi^+
({\vec\sigma}_N\cdot{\vec\tau}_1)\xi$ corresponds to the absorption of the virtual
pion by the final nucleon (the bottom $\pi N$ vertex in Fig.2b) and the vertex
$\Gamma_N^{(2)}={\bar u}(k_1^f)\gamma_5 u(p)=\xi^+{\vec\sigma}_N\cdot{\vec\tau}_2\xi$ 
corresponds to the emission of the virtual pion by initial nucleon (the top $\pi-N$
vertex in Fig.2), here $u$ is the four-component spinor of the nucleon, 
whereas ${\bar u}$ is the conjugated four-component spinor of the nucleon in the
deuteron; $\xi$ is the two-component spinor of the nucleon; the forms for the vectors
${\vec \tau}_1,{\vec \tau}_2$ are presented in the APPENDIX.
The amplitude of the elastic $\pi-N$ scattering $f_{\pi N}(s_1,t_1)$ depends on the
square of the energy in the $\pi-N$ c.m.s. $s_1=(q_2+k_2)^2$ and the
the four-momentum transferred square $t_1=(q_2-q_1)^2$, where $q_2,q_1$ are the
four-momenta of the virtual pion before and after the $\pi-N$ scattering, $k_1,k_2$
are the four-momenta of proton and neutron in the initial deuteron with the
four-momentum $p_d$, whereas $k_1^\prime,k_2\prime$ are the four-momenta of these
nucleons in the final deuteron with the four-momentum $p_d^\prime$. In the $pd$
c.m.s. the three-momenta of nucleons in the initial and final deuteron can be
presented in the following form:
\be
{\vec k}_1~=~\frac{1}{2}{\vec p}_d-{\vec k},~{\vec k}_2~=~\frac{1}{2}
%{\vec p}_d+{\vec k};~
{\vec p}_d+{\vec k};\\
%\nonumber
{\vec k}_1^f~=~\frac{1}{2}{\vec p}_d^ f-{\vec k}^\prime,~{\vec k}_2^f~
=~\frac{1}{2}
{\vec p}_d^f+{\vec k}^ \prime~,
\label{def:vectorsk}
\ee 
where ${\vec k}$ and ${\vec k}^\prime$ are the relative momenta of nucleons in the
initial and final deuterons respectively. 

To calculate the amplitude ${\cal F}^{(3)}$ given by Eq.(\ref{def:GOPT}), we 
removed the integral $f_{\pi N}$ at the mean value of the nucleon relative momentum
in deuteron ${\bar \mid{\vec k}\mid}\simeq  0.07-0.1$ GeV$/c$ because the d.w.f.
$\Psi_d(x,{\vec k}_t)$ is sharply decreasing as $x$ and $\mid{\vec k}_t\mid$ grow,
as it was done, for example, in 
\cite{DL:1990,Kolybasov:1971}.
The amplitude $f_{\pi N}^{el}$ was presented in the following form 
\cite{Ponomarev:1976}:  
\begin{equation}
f_{\pi N}^{el}(s_1,t_1)~=~A(s_1,t_1)~+~iB(s_1,t_1)({\vec\sigma}\cdot{\vec n})~,
\label{def:fpiN}
\end{equation}
where $n=({\vec q}_2^*\times {\vec q}_1^*)/\mid({\vec q}_2^*\times {\vec q}_1^*)\mid$
is the unit vector, ${\vec q}_2^*$ and ${\vec q}_1^*$ are the three-momenta of the
intermediate pion before and after $\pi-N$ scattering in the $\pi-N$ c.m.s. 
The details for the kinematics corresponding to the elastic $\pi-N$ scattering and the 
backward $d p$ scattering are presented in the APPENDIX. The functions $A(s_1,t_1)$ and 
$B(s_1,t_1)$ were found from the phase shift analysis for the elastic $\pi N$
scattering
\cite{Strakovsky:2003}.  

%%%%%%%%%%%%%%%%%%%%%%%%%%%%%%%%%%%%%%%%%%%%%%%%%%%%%%%%%%%%%%%%%%%%%%%%%%%%%%%
\subsection{Observables for $dp\rightarrow pd$ reaction }
We calculated the differential cross section $d\sigma/d\Omega$, the transfer
polarization $\kappa$ and the tensor analyzing power $T_{20}$.
\begin{equation}
\frac{d\sigma}{d\Omega}~=~\frac{1}{64\pi^2 s}\mid{\cal F}^{tot}\mid^2~,
\label{def:diffcrs}
\end{equation}
where the total amplitude ${\cal F}^{tot}$ calculated, for example, within the 
LCD has the following form:
\begin{equation}
{\cal F}^{tot}_{LCD}~=~{\cal F}^{(1)}_{LCD}~+~{\cal F}^{(3)}_{LCD}~.
\label{def:FtotLCD}
\end{equation}
Assuming that the calculation of the triangle graphs of Fig.3 within the LCD 
can give the same results as the calculation of the
triangle Feynman graph of Fig.2 we also compute the sum of the Feynman graph
of Fig.2b and the diagram of Fig.3c. The total amplitude within this combined 
relativistic calculation (RC) is presented in the following form:
\begin{equation}
{\cal F}^{tot}_{RC}~=~{\cal F}^{(RONE)}~+~{\cal F}^{(3)}_{LCD}~.
\label{def:FtotRC}
\end{equation}  
Then we compare all the results obtained within the LCD using the total amplitude given 
by Eq.(\ref{def:FtotLCD}) and the RC using Eq.(\ref{def:FtotRC}) for ${\cal F}^{tot}$. 
The reason for this assumption is based on the results of \cite{Brodsky:1973} which show that
the triangle diagrams with $x<0$ can be more suppressed than the $Z$-diagrams within the impulse
approximation of Fig.1c. 

The tensor analyzing power of the deuteron $T_{20}$ has the following form:
\begin{equation}
T_{20}~=~\frac{Tr\left(\rho_d({\cal F}^{tot})^+\Omega_{20}{\cal F}^{tot}\right)}
{Tr\left(\rho_d({\cal F}^{tot})^+{\cal F}^{tot}\right)}~,
\label{def:T20}
\end{equation}
where \cite{Vasan:1973}
\begin{equation}
\!\Omega_{20}~=~\frac{1}{\sqrt{2}}\left(3S_z^2-2\right)\equiv
\frac{1}{\sqrt{2}}\left(\frac{3}{2}(1+\sigma_{pz}\sigma_{nz})-2\right)\!
\label{def:Omega20}
\end{equation}
is the spin-tensor operator corresponding to the tensor component of the deuteron
polarization. Here $S_z$ is the projection of the deuteron spin operator $S$ on the 
quantization axis $z$, which in our case is the direction of the initial deuteron,
whereas $\sigma_{pz}$ and $\sigma_{nz}$ are the $z$ components of the Pauli matrices
corresponding to the proton and the neutron, respectively.
The transfer polarization and the tensor analyzing power for the deuteron
were studied within the impulse approximations in
\cite{Ableev:1988,Strokovsky:1999}.
The transfer polarization is defined as 
\begin{equation}
\kappa_0~=~\frac{{\vec{\cal P}}^\prime\cdot {\vec n}}{{\vec{\cal P}}\cdot {\vec n}
(1-\rho_{20}T_{20})}~,
\label{def:kappa}
\end{equation}
where ${\vec{\cal P}}^\prime$ is the vector polarization of the final proton,
${\vec n}$ is the unit vector transverse to the reaction plane, ${\vec{\cal P}}$ 
is the vector polarization of the initial deuteron and
\begin{equation}
{\vec{\cal P}}^\prime\cdot {\vec n}~=~\frac{Tr\left(\rho_d({\cal F}^{tot})^+
{\vec\sigma}\cdot{\vec n}{\cal F}^{tot}\right)}
{Tr\left(\rho_d({\cal F}^{tot})^+{\cal F}^{tot}\right)}~,
\label{def:Pprimen}
\end{equation}
$\rho_d$ is the density matrix of the deuteron, it has the following form:
\begin{equation}
\rho_d~=~\frac{1}{3}P_T\left(1+\frac{3}{2}{\vec{\cal P}}\cdot {\vec S}-
\frac{1}{2}\rho_{20}(3S_z^2-2)\right)~.
\label{def:rhod}
\end{equation}
Here $P_T=(3+{\vec\sigma}_p\cdot {\vec\sigma}_n)/4$ is the projection operator of
the triplet deuteron state, $\rho_{20}$ is its tensor polarization.

Calculating the traces in Eqs.
(\ref{def:Pprimen},\ref{def:T20})
we have the following general form for $\kappa$ and $T_{20}$:
\begin{equation}
\kappa_0~=~\frac{u^2(k^2)-\frac{1}{\sqrt{2}}u(k^2)w(k^2)-w^2(k^2)+
{\tilde\Delta}_\kappa}
{(u^2(k^2)+w^2(k^2)+\Delta)(1-\rho_{20}T_{20}^{sp})}~,
\label{def:kappatot}
\end{equation}
\begin{equation}
T_{20}~=~\frac{1}{\sqrt{2}}\frac{2\sqrt{2}u(k^2)w(k^2)-w^2(k^2)+
{\tilde\Delta}_{T_{20}}}{u^2(k^2)+w^2(k^2)+\Delta}~,
\label{def:T20tot}
\end{equation}
where ${\tilde\Delta}_\kappa,{\tilde\Delta}_{T_{20}}$ and $\Delta$ are the
corrections due to the  contributions of the triangle graphs. Here $T_{20}^{sp}$ is
the tensor analyzing power of the deuteron calculated within the spectator model
\cite{Ableev:1988}
that has the form given by Eq.
(\ref{def:T20tot})
at ${\tilde\Delta}_{T_{20}}=\Delta=0$. In the spectator model, when
${\tilde\Delta}_\kappa=\Delta=0$, the transfer polarization was analyzed in
details in
\cite{Strokovsky:1990,Strokovsky:1999}.
The forms for the correction functions ${\tilde\Delta}_\kappa,\Delta$ and 
${\tilde\Delta}_{T_{20}}$ are presented in the APPENDIX.

%%%%%%%%%%%%%%%%%%%%%%%%%%%%%%%%%%%%%%%%%%%%%%%%%%%%%%%%%%%
\section{Results and discussion}
We calculated the center-of-mass differential cross section $d\sigma/d\Omega$, 
the tensor analyzing power of the deuteron $T_{20}$ and the transfer polarization 
$\kappa_0$ in the elastic backward $p-d$ scattering.
This calculation was done within the RONE (the graph of Fig.1a) 
and the impulse approximation of the LCD (the graph of Fig.1b).We also calculated these
observables including both simple graphs of Fig.1 and the triangle graphs of Fig.3. 
These results obtained within the LCD and 
the RC are presented in Figs.(4,5) as a function of the deuteron momentum $p_d^{l.s.}$ 
in the laboratory system (l.s.).
In Fig.4 curves 1 and 2 correspond to the total calculation within the RC,
see Eq.(\ref{def:FtotRC}) for ${\cal F}^{tot}_{RC}$ with the Reid soft core
d.w.f. \cite{Reid:1968} and the  Argon-18 d.w.f. \cite{AV18} respectively;
curves 3 and 4 correspond to the LCD, see Eq.(\ref{def:FtotLCD}) for 
${\cal F}^{tot}_{LCD}$ with the same kinds of the d.w.f.; curves 5 and 6 correspond 
to the RONE (Fig.1a) and the LCD impulse approximation (Fig.1b) with the Reid soft 
core d.w.f. \cite{Reid:1968} and curves 7,8 correspond to the same calculations as 
for curves 5,6 but with the AV18 d.w.f. \cite{AV18}.
One can see from Fig.4 that the total calculation within the LCD and RC using
both the Reid soft cor{e d.w.f. and the AV18 d.w.f. give approximately the same 
results for the differential cross section which are very close to the experimental data
that are taken from \cite{Azhgirei}. As is seen from Fig.4 both impulse
approximations corresponding to Fig.1a and Fig.1b do not describe the experimental
data on $d\sigma/d\Omega$ at $p_d^{l.s.}~>~1.5$ GeV$/c$. 
\begin{center}
\begin{figure}[htb]
\includegraphics[width=0.45\textwidth]{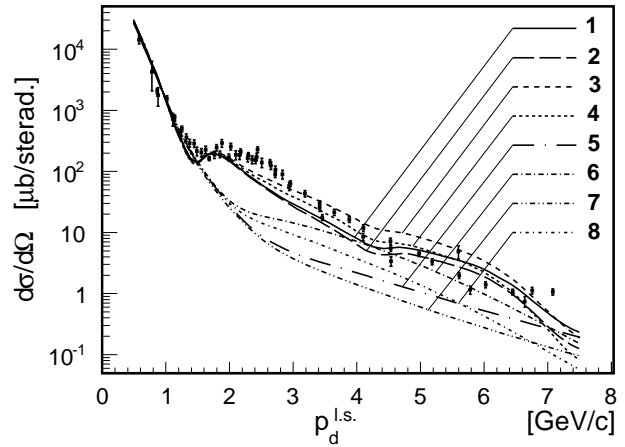}
\caption[Fig.4]{The center-of-mass differential cross section  
$d\sigma/d\Omega_{c.m.s.}$ for the elastic backward $p-D$ scattering  
as a function of the deuteron momentum $p_d^{l.s}$ in the laboratory system.} 
\end{figure}
\end{center}
\begin{figure}[htb]
{\includegraphics[width=0.45\textwidth]{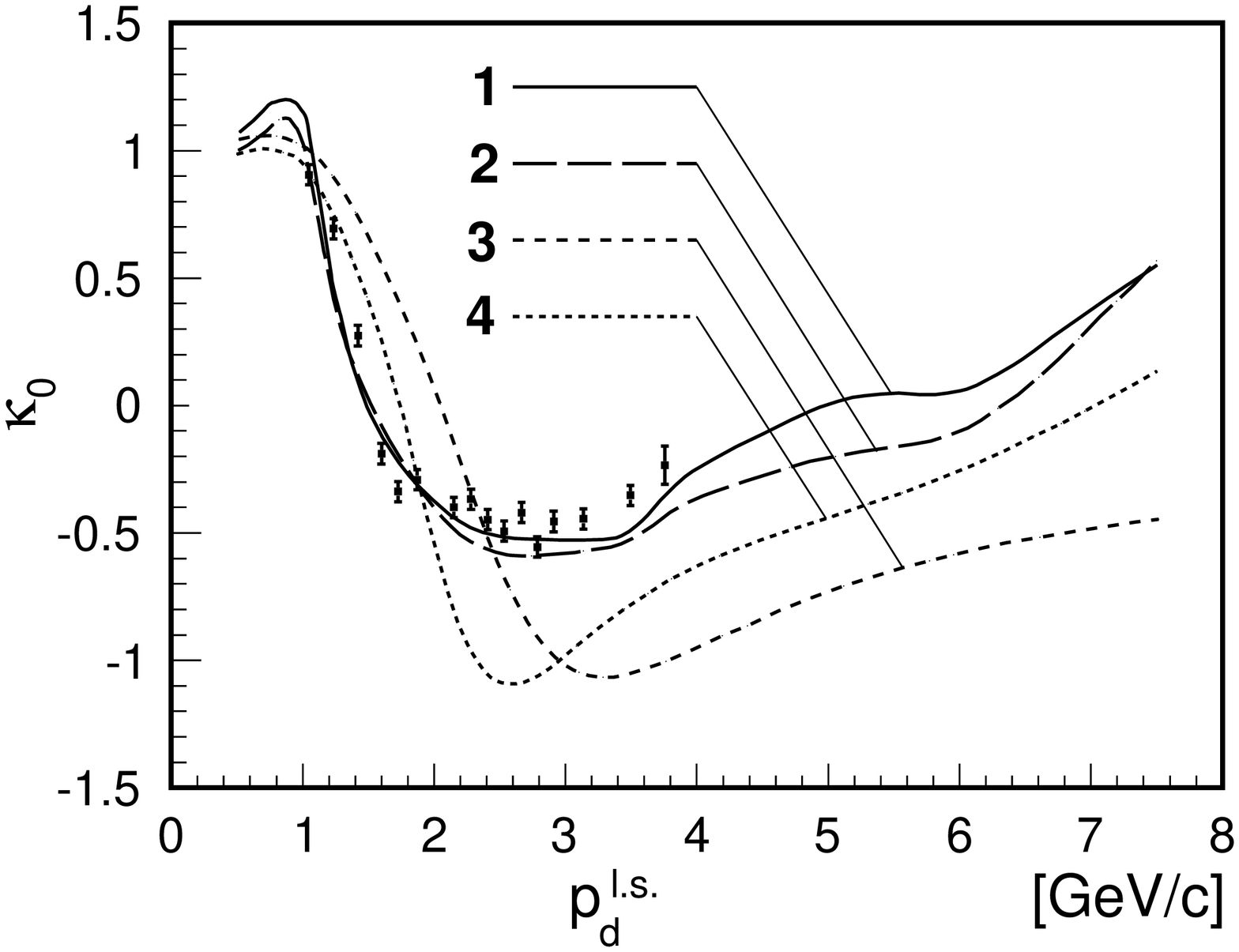}}
{\includegraphics[width=0.45\textwidth]{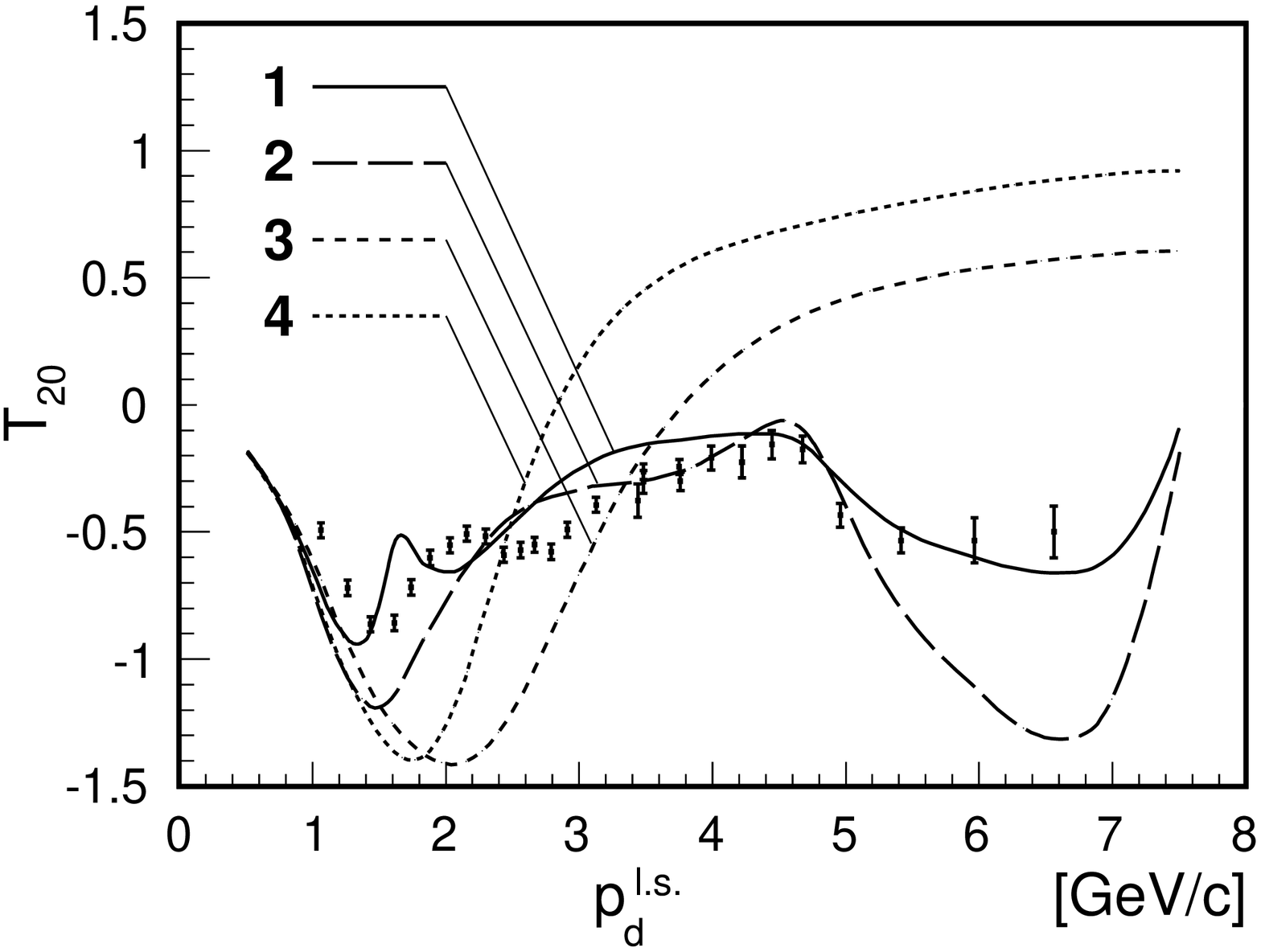}}
\caption[Fig.5]{The tensor analyzing power of the deuteron $T_{20}$ as a function
of $p_d^{l.s}$ (top)
and the transfer polarization $\kappa_0$ as a function of $p_d^{l.s}$ (bottom).}
\end{figure}

In Fig.5 the tensor analyzing power of the deuteron $T_{20}$ (top ) and the 
transfer polarization $\kappa$ (bottom) are presented as a function of the 
initial deuteron momentum $p_d$ in the l.s. using the Reid 
soft core $N-N$ potential for the d.w.f.\cite{Reid:1968}. 
Curves 1 and 2 in Fig.5 correspond to the total calculation within the RC and the LCD 
respectively, whereas curves 4 and 5 correspond to the RONE calculation (Fig.1a) and 
the LCD impulse approximation (Fig.1b). One can see from Fig.5 that the impulse 
approximations (Fig.1a and Fig.1b) do not describe $T_{20}$ and $\kappa_0$ at
$p_d^{l.s.}~>~1$ GeV$/c$. As is seen from Fig.5 (bottom), the total calculations of 
the transfer polarization $\kappa_0$ within both the RC and the LCD give the same 
results at $p_d^{l.s.}~\leq~4$ GeV$/c$ which are very close to the experimental data.
The not so large difference between the RC and the LCD calculations of $\kappa_0$
appears at $p_d^{l.s.}~>~4$ GeV$/c$, where no experimental data are available now.   
Therefore, analyzing the experimental data on the transfer polarization
one can not differentiate between the RC and the LCD calculations.
In contrast, Fig.5 (top) shows that the tensor analyzing power $T_{20}$ is very 
sensitive to the total calculations within the RC and the LCD approximations.
As is seen from Fig.5 (top), the total RC calculation results in a better description
of the experimental data on $T_{20}$ in the whole region of the initial deuteron momenta, whereas
the LCD calculation gives a worse description of the data at $1.2~<~p_d^{l.s.}~<~1.8$ GeV$/c$
and especially at $p_d^{l.s.}~>~5$ GeV$/c$. It can be due to a sizable contribution from the
$Z$-diagram of Fig.1c to $T_{20}$ that is included by the Feynman graph of Fig.1a corresponding 
to the relativistic one nucleon exchange (RONE). 
On the other hand, as is mentioned above, the inclusion of the relativistic triangle Feynman
graph of Fig.2a in \cite{Nakamura:1985} did not allow a description of $T_{20}$ at 
$p_d^{l.s.}~>~4$ GeV$/c$
that corresponds to the intradeuteron nucleon momenta $k~>~0.5$ GeV$/c$ or the light cone
variables $x~>~0.4$ \cite{GL:1993}. It can probably be caused by the following. In the calculation of
the Feynman graph of Fig.2a the relativistic invariant $d-N$ vertex is related in \cite{Nakamura:1985} 
to the nonrelativistic d.w.f., while within the LCD we relate the $d-N$ vertex to $\Phi^{n.r.}_d(k^2)$
for the time-ordered graphs corresponding only to the deuteron dissociation (Fig.3(a,b)) and neglect
the $Z$-diagrams in the triangle graphs corresponding to the annihilation ${\bar N}d\rightarrow N$.
This is the difference between our calculation of the triangle diagram within the LCD (Fig.3) and
the calculation of the Feynman triangle graph (Fig.2a) in \cite{Nakamura:1985}.

Note that we also calculated all the observables presented in Figs.(4,5) using the CD Bonn d.w.f.
\cite{CDBonn} and the N3L0 d.w.f. \cite{N3LO}; however, the description of $T_{20}$ and
$\kappa_0$ was worse, especially at $p_d^{l.s.}>2$ GeV$/c$. Therefore, we do not present these
results because the figures will be rather cumbersome. Nevertheless, in the APPENDIX we present 
the approximations of these d.w.f. by the simple Gauss forms that could be useful for other 
calculations.  
%%%%%%%%%%%%%%%%%%%%%%%%%%%%%%%%%%%%%%%%%%%%%%%%%%%%%%%%%%%%%%%%%%%%%%%%%%%%%%%%%%%%%%

\section{ Conclusion}

The theoretical analysis of the elastic backward $p-d$ or the forward $d-p$ scattering 
within the light cone dynamics allows us to draw the following conclusions. The calculation 
of the differential cross section and the polarization observables, the tensor analyzing 
power $T_{20}$ and the transfer polarization $\kappa_0$ within the impulse approximation 
(diagrams of Fig.1) is not able to describe the experimental data at the initial deuteron
momenta $p_d\geq 1.2 $ GeV$/c$. In this kinematic region the contribution 
of the triangle graphs (Fig.3) is very sizable because it is mainly due to the
possible creation of the $\Delta$ isobar in the intermediate state. The inclusion
of these graphs results in a rather satisfactory description of the experimental data
on the differential cross section $d\sigma/d\Omega$ in a wide region of the
initial momenta. We show that the contribution of the RONE graph (Fig.1a)
and the LCD impulse approximation (Fig.1b) give approximately similar results for the differential cross 
section; therefore, the contribution of the $Z$-diagram
(Fig.1c) to $d\sigma/d\Omega$ is not large. However, its contribution to the tensor analyzing
power is sizable because the total RC calculation including graphs of Fig.1a and Fig.3 
describes the experimental data on $T_{20}$ better than the total LCD calculations
(graphs of Fig.1b and Fig.3). The experimental data on the transfer polarization $\kappa_0$
are described rather satisfactorily by both the total RC and total LCD calculations. 
One can conclude that the calculation of all the observables for the elastic
backward $p-d$ scattering within the light cone dynamics including the triangle graphs
of Fig.3 and the $Z$-diagrams of Fig.1c results in a rather satisfactory description
of the experimental data at initial deuteron momenta up to $7 GeV/c$. Note that we
do not include the six-quark admixture in the deuteron wave function. 
This effect can probably be important at larger  initial momenta because the
contribution of the graphs of Figs.(1-3) decreases with increasing $p_d^{l.s}$, as is shown
in Fig.4.   
%%%-----------  Fig.4 1-pos.----------------
%%%-----------  Fig.5 1-pos.----------------
%%%%%%%%%%%%%%%%%%%%%%%%%%%%%%%%%%%%%%%%%%%%%%%%%%%%%%%%%%%%%%%%%%%%%%%%%%%%%%%%%

{\bf  Acknowledgment.}
We are very grateful to E.A.Strokovsky for extremely useful discussions and
help in the preparation of this paper. We also thank 
 F.Gross, V.A.Karmanov, A.P.Kobushkin, I.M.Sitnik and Yu.N.Uzikov
for very useful discussions. This work was supported in part by the RFBR grant 
No. 08-02-01003.
%\bibliography{paper_jl}

%%%%%%%%%%%%%%%%%%%%%%%%%%%%%%%%%%%%%%%%%%%%%%%%%%%%%%%%%%%%%%%%%%%%%%%%%%%%%%%%%
\begin{widetext}

\section{\bf Appendix}
%\begin{widetext}
\subsection{\bf Corrections ${\tilde\Delta}_{T_{20}},{\tilde\Delta}_{\kappa_0},\Delta $}
%\begin{widetext}
 \vspace{1.0cm}
Let us present the general forms for the corrections ${\tilde\Delta}_{\kappa_0}$,
${\tilde\Delta}_{T_{20}}$ and $\Delta$ entering into  Eq.(\ref{def:T20}) for $T_{20}$
and Eq.(\ref{def:kappa}) for $\kappa_0$.
\begin{equation}
{\tilde\Delta}_{T_{20}}~=~Tr\left(\rho_d[({\cal F}^{(3)})^+
\Omega_{20}{\cal F}^{(1)}+({\cal F}^{(1)})^+\Omega_{20}{\cal F}^{(3)}+
({\cal F}^{(3)})^+\Omega_{20}{\cal F}^{(3)}]
\right)
\label{def:DeltaT20}
\end{equation}
\vspace{0.3cm}
\begin{equation}
{\tilde\Delta}_{\kappa_0}~=~Tr\left(\rho_d[({\cal F}^{(3)})^+
{\vec\sigma}\cdot{\vec n}{\cal F}^{(1)}+
({\cal F}^{(1)})^+{\vec\sigma}\cdot{\vec n}{\cal F}^{(3)}+
({\cal F}^{(3)})^+{\vec\sigma}\cdot{\vec n}{\cal F}^{(3)}]
\right)
\label{def:Deltakappa}
\end{equation}
 \vspace{0.3cm}
\begin{equation}
\Delta~=~Tr\left(\rho_d[2Re(({\cal F}^{(3)})^+
{\cal F}^{(1)})+({\cal F}^{(3)})^+{\cal F}^{(3)}]\right)
\label{def:Delta}
\end{equation}
\end{widetext}
\subsection{\bf Vertices $\Gamma_N^{(1)},\Gamma_N^{(2)}$}
The vector $ {\vec\tau}_1$ entering into the $\pi$-absorption vertex  
\begin{equation}
\Gamma_N^{(1)}={\bar u}(p^f)\gamma_5 u(k_1)=\xi^+({\vec\sigma}_N\cdot{\bf \tau}_1)\xi 
\label{def:GamNf}
\end{equation}
has the following form:
\begin{equation}
{\vec\tau}_1=a_1\frac{{\vec p}^f-{\vec k}_1}{2m}-b_1\frac{E_N({\vec p}^f)-E_N({\vec k}_1)}
{2m}\frac{{\vec p}^f+{\vec k}_1}{2m}~,
\label{def:tauf}
\end{equation}
where
\begin{equation}
a_1=\frac{(E_N({\vec p}^f)+E_N({\vec k}_1))/2+m}
{\sqrt{(E_N({\vec p}^f)+m)(E_N({\vec k}_1)+m)}}~,
\label{def:afirst}
\end{equation}
\begin{equation}
b_1=\frac{1}
{\sqrt{(E_N({\vec p}^f)+m)(E_N({\vec k}_1)+m)}}~.
\label{def:bfirst}
\end{equation}
The vector ${\vec\tau}_2$ entering into the $\pi$-emission vertex  
\begin{equation}
\Gamma_N^{(2)}={\bar u}(k_1^f)\gamma_5 u(p)=\xi^+{\vec\sigma}_N\cdot{\vec \tau}_2\xi
\label{def:GamNs}
\end{equation}
has the form
\begin{equation}
{\vec\tau}_2=a_2\frac{{\vec k}^f_1-{\vec p}}{2m}-b_2\frac{E_N({\vec k}^f_1)-E_N({\vec p})}
{2m}\frac{{\vec k}^f_1+{\vec p}}{2m}~,
\label{def:taus}
\end{equation}
where 
\begin{equation}
a_2=\frac{(E_N({\vec p})+E_N({\vec k}^f_1))/2+m}
{\sqrt{(E_N({\vec p})+m)(E_N({\vec k}^f_1)+m)}}
\label{def:asecond}
\end{equation}
and
\begin{equation}
b_2=\frac{1}
{\sqrt{(E_N({\vec p})+m)(E_N({\vec k}^f_1)+m)}}~.
\label{def:bsecond}
\end{equation}

\subsection{\bf Kinematics for elastic $\pi-N$ and backward $p-d$ scattering }
The square of the initial energy in the $\pi-N$ c.m.s. reads
\begin{equation}
s_1=(q_2+k_2)^2~,
\end{equation}
where $q_2$ and $k_2$ are the four-momenta of the colliding intermediate pion and a nucleon 
in the initial deuteron. Introducing the variable ${\vec\Delta}={\vec p}_d^f/2-{\vec p}$
and using Eqs.(\ref{def:vectorsk}) one can get the following form for $s_1$:
\begin{equation}
s_1\simeq\left(\mid{{\vec k}^\prime-{\vec\Delta}}\mid+E_N(p_d/2)\right)^2-
({\vec k}^\prime-{\vec\Delta})\cdot ({\vec k}+{\vec p}_d/2)~,
\label{def:sfirst}
\end{equation}
The transfer in the $\pi-N$ elastic scattering is 
\begin{equation}
t_1=(q_2-q_1)^2~,
\end{equation}
where $q_1$ is the four-momentum of the rescattered pion. Introducing the variable
${\vec\Delta}^\prime={\vec p}_d/2-{\vec p}^f$ and taking into account that for the 
backward $p-d$ scattering ${\vec\Delta}^\prime=-{\vec\Delta}$ we have the following form for
$t_1$:
\begin{equation}
t_1\simeq\left(\mid{{\vec k}^\prime-{\vec\Delta}}\mid- 
\mid{{\vec k}+{\vec\Delta}}\mid\right)^2-({\vec k}^\prime-{\vec k}-2{\vec\Delta})^2~.
\label{def:tfirst}
\end{equation}
Note that getting Eqs.(\ref{def:sfirst},\ref{def:tfirst}) we neglected the pion mass squared 
$\mu_\pi^2$.
%%%%%%%%%%%%%%%%%%%%%%%%%%%%%%%%%%%%%%%%%%%%%%%%%%%%%%%%%%%%%%%%%%%
\subsection{\bf Deuteron wave functions}
 \vspace{0.2cm}
We presented the d.w.f of the type of Reid soft core \cite{Reid:1968},  AV18 \cite{AV18},
 N3LO \cite{N3LO} and  CD Bonn \cite{CDBonn} in the following forms of the Gauss functions
and found all the parameters from their fits. 
  
\begin{equation}
u(p)~=~\sum_{n=1}^{n_{max}}A_n\exp(-\alpha_n p^2)
\label{def:up}
\end{equation}
 \vspace{-0.5cm}
\begin{equation}
w(p)~=~p^2\sum_{n=1}^{n_{max}}B_n\exp(-\beta_n p^2)
\label{def:wp}
\end{equation}

%\end{widetext}

 \vspace{0.2cm}  %-1.0cm
\begin{table}[h]
\caption{\bf The Reid soft core d.w.f.~\cite{Reid:1968}  ($n_{max}=5$):}
%{\baselineskip  6pt%18pt
 \vspace{0.2cm}
\label{Tb1}
\begin{center}
\begin{tabular}[t]{|c|c|c|c|c|}
\hline
 $n$ & $A_n$ & $\alpha_n$ & $B_n$ & $\beta_n$ \\
\hline
 1 & 9.007 & 1277.26 & 1.358 & 5.165 \\
\hline
 2 & 20.035 & 370.595 & 11.289 & 15.774 \\
\hline
 3 & 9.724 & 88.625 & 15.376 & 50.065 \\
\hline
 4 & 2.142 & 18.904 & 43.963 & 52.592 \\
\hline
 5 & -0.184 & 2.494 & 227.617 & 205.697 \\
\hline
\end{tabular}
\end{center}
%}
\end{table}

 \vspace{0.5cm} %-1.0cm
\begin{table}[h]
\caption{\bf AV18 d.w.f.~\cite{AV18}  ($n_{max}=7$):}
%{\baselineskip 6pt%18pt
% \vspace{-0.1cm}
\vspace{0.1cm}
\label{Tb2}
\begin{center}
\begin{tabular}[t]{|c|c|c|c|c|}
\hline
 $n$ & $A_n$ & $\alpha_n$ & $B_n$ & $\beta_n$ \\
\hline
 1 & 5.33818388 & 1277.26 & 1.26183415 & 5.165 \\
\hline
 2 & 17.7506951 & 370.595 & 10.7790333 & 15.774 \\
\hline
 3 & 10.1156672 & 88.625 & -30.4158329 & 50.065 \\
\hline
 4 & 2.00231994 & 18.904 & 91.7607541 & 52.592 \\
\hline
 5 & -0.129987968 & 2.494 & 193.350066 & 205.697 \\
\hline
 6 & 1.85353863 & 15000.0 & 51.1855721 & 600.0 \\
\hline
 7 & 4.94736493 & 650.0 & 221.427665 & 1000.0 \\
\hline
\end{tabular}
\end{center}
%}
\end{table}

 \vspace{0.5cm}
\begin{table}[h]
\caption{\bf CD Bonn d.w.f.~\cite{CDBonn}  ($n_{max}=7$):}
%{\baselineskip 6pt%18pt
 \vspace{0.5cm}
 %\vspace{-0.5cm}
\label{Tb3}
\begin{center}
\begin{tabular}[t]{|c|c|c|c|c|}
\hline
 $n$ & $A_n$ & $\alpha_n$ & $B_n$ & $\beta_n$ \\
\hline
 1 & 6.25524152 & 1277.26 & 0.836559861 & 5.165 \\
\hline
 2 & 18.0668014 & 370.595 & 11.225581 & 15.774 \\
\hline
 3 & 10.0949093 & 88.625 & -54.150829 & 50.065 \\
\hline
 4 & 1.96622794 & 18.904 & 117.382885 & 52.592 \\
\hline
 5 & -0.0681124745 & 2.494 & 192.905072 & 205.697 \\
\hline
 6 & 1.02009756 & 15000.0 & 83.961456 & 600.0 \\
\hline
 7 & 4.06853301 & 650.0 & 161.799081 & 1000.0 \\
\hline
\end{tabular}
\end{center}
%}
\end{table}

 \vspace{0.5cm}  %-1.0cm
\begin{table}[h]
\caption{\bf N3LO d.w.f.~\cite{N3LO}  ($n_{max}=7$):}
%{\baselineskip 6pt%18pt
 \vspace{-0.1cm}
\label{Tb4}
\begin{center}
\begin{tabular}[t]{|c|c|c|c|c|}
\hline
 $n$ & $A_n$ & $\alpha_n$ & $B_n$ & $\beta_n$ \\
\hline
 1 & 5.75328843 & 1277.26 & -0.411245062 & 5.165 \\
\hline
 2 & 17.6314 & 370.595 & 17.5832912 & 15.774 \\
\hline
 3 & 10.125005 & 88.625& -251.958128 & 50.065 \\
\hline
 4 & 2.1933269 & 18.904 & 318.379763 & 52.592 \\
\hline
 5 & -0.228481595 & 2.494 & 166.840281 & 205.697 \\
\hline
 6 & 0.787045975 & 15000.0 & 194.835805 & 600.0 \\
\hline
 7 & 4.82444602 & 650.0 & 3.08550887 & 1000.0 \\
\hline
\end{tabular}
\end{center}
%}
\end{table}

\vspace{10.5cm}
\newpage
\bibliography{paper_jl_prc}
%\bibliography{paper2_jl}

\end{document}